\documentclass[aps, prl, reprint,superscriptaddress,showpacs,preprintnumbers,amsmath,amssymb, hidelinks, nofootinbib]{revtex4-1}

\pdfoutput=1

\usepackage{amsmath}
\usepackage{graphicx}
\usepackage{natbib}
\usepackage{verbatim}
\usepackage[withbib, all]{authorindex} 
\usepackage[usenames]{color}
\usepackage{bm}
\usepackage{hyperref}
\usepackage{xcolor}
\usepackage[english]{babel}
\usepackage{braket}

\hypersetup{
colorlinks, linkcolor={blue},
citecolor={blue}, urlcolor={blue},
}

\begin{document}

\title{Optically Detected Ferromagnetic Resonance in Metallic Ferromagnets via Nitrogen Vacancy Centers in Diamond}

\author{M. R. Page*}
\affiliation{Department of Physics, The Ohio State University, Columbus, Ohio 43210, USA}

\author{F. Guo*}
\affiliation{School of Applied and Engineering Physics, Cornell University, Ithaca NY 14853}

\let\thefootnote\relax\footnote{* M. R. Page and F. Guo contributed equally to this work.}

\author{C. M. Purser}
\affiliation{Department of Physics, The Ohio State University, Columbus, Ohio 43210, USA}

\author{J. G. Schulze}
\affiliation{Department of Physics, The Ohio State University, Columbus, Ohio 43210, USA}

\author{T. M. Nakatani}
\affiliation{San Jose Research Center, HGST, a Western Digital company, San Jose, California 95135, USA}

\author{C. S. Wolfe}
\affiliation{Department of Physics, The Ohio State University, Columbus, Ohio 43210, USA}

\author{J. R. Childress}
\affiliation{San Jose Research Center, HGST, a Western Digital company, San Jose, California 95135, USA}

\author{P. C. Hammel}\email{hammel@physics.osu.edu}
\affiliation{Department of Physics, The Ohio State University, Columbus, Ohio 43210, USA}

\author{G. D. Fuchs}\email{gdf9@cornell.edu}
\affiliation{School of Applied and Engineering Physics, Cornell University, Ithaca NY 14850}

\author{V. P. Bhallamudi}
\affiliation{Department of Physics, The Ohio State University, Columbus, Ohio 43210, USA}

\date{\today}

\begin{abstract}
We report quantitative measurements of optically detected ferromagnetic resonance (ODFMR) of ferromagnetic thin films that use nitrogen-vacancy (NV) centers in diamonds to transduce FMR into a fluorescence intensity variation. To uncover the mechanism responsible for these signals, we study ODFMR as we 1) vary the separation of the NV centers from the ferromagnet (FM), 2) record the NV center longitudinal relaxation time $T_1$ during FMR, and 3) vary the material properties of the FM. Based on the results, we propose the following mechanism for ODFMR. Decay and scattering of the driven, uniform FMR mode results in spinwaves that produce fluctuating dipolar fields in a spectrum of frequencies. When the spinwave spectrum overlaps the NV center ground-state spin resonance frequencies, the dipolar fields from these resonant spinwaves relax the NV center spins, resulting in an ODFMR signal. These results lay the foundation for an approach to NV center spin relaxometry to study FM dynamics without the constraint of directly matching the NV center spin-transition frequency to the magnetic system of interest, thus enabling an alternate modality for scanned-probe magnetic microscopy that can sense ferromagnetic resonance with nanoscale resolution.
\end{abstract}

\maketitle


Understanding magnetic dynamics in future storage and information processing technologies will be a key to their development \citep{bader_spintronics_2010,zutic_spintronics:_2004}. In particular, it will be necessary to measure and understand relaxation \citep{tserkovnyak_enhanced_2002,slonczewski_current-driven_1996}, angular momentum transfer \citep{ralph_spin_2008,qiu_spin_2013,adur_damping_2014,adur_microscopic_2015} and spinwave propagation \citep{vashkovskii_propagation_1996,kajiwara_transmission_2010,demokritov_spin_2009}, not only in extended magnetic films, but also in nanoscale devices \citep{krivoruchko_spin_2015}. In addition, establishing new mechanisms for imaging magnetization dynamics in confined structures will aid in improving current magnetic technologies \citep{childress_all-metal_2008, parkin_exchange-biased_1999,ikeda_perpendicular-anisotropy_2010, parkin_magnetic_2008} and enhance them using emerging materials such as those featuring magnetic textures \citep{nagaosa_topological_2013,tokunaga_new_2015,fert_skyrmions_2013}. 

The nitrogen-vacancy (NV) center in diamond has emerged as a flexible and sensitive platform for nanoscale magnetic sensing \citep{schirhagl_nitrogen-vacancy_2014, balasubramanian_nanoscale_2008, maze_nanoscale_2008} due to its atomic-scale size and its spin-sensitive fluorescence, enabling optical detection of magnetic dynamics \citep{tetienne_nanoscale_2014,van_der_sar_nanometre-scale_2015,dussaux_observation_2015}. NV-based magnetometry aimed at dynamic magnetic fields have typically required either spin-echo protocols \citep{taylor_high-sensitivity_2008}, which are constrained to frequencies that are quasi-static compared to FMR (e.g. $\sim$ MHz or below), or it requires direct resonance with an NV center spin transition \citep{Wang_NatComm_2015, Appel_NJP_2016}. 

In contrast, we have recently demonstrated an alternate modality \citep{wolfe_off-resonant_2014,wolfe_spatially_2016} for detecting ferromagnetic resonance with diamond NV centers placed in nanoscale proximity to Yttrium Iron Garnet (YIG) that uses a simple, continuous wave (CW) protocol. A surprising observation was the change of NV center fluorescence due to ferromagnetic resonance at frequencies that were well separated from the NV center spin resonance.  Conveniently, these signals were acquired with no direct resonant manipulation of the NV center spins, making them ideal for integration with future NV center-based scanned-probe microscopy of magnetic resonance.  Establishing the mechanism of these signals is key to their future use.

\begin{figure}
\center{\includegraphics[width=0.95\linewidth]{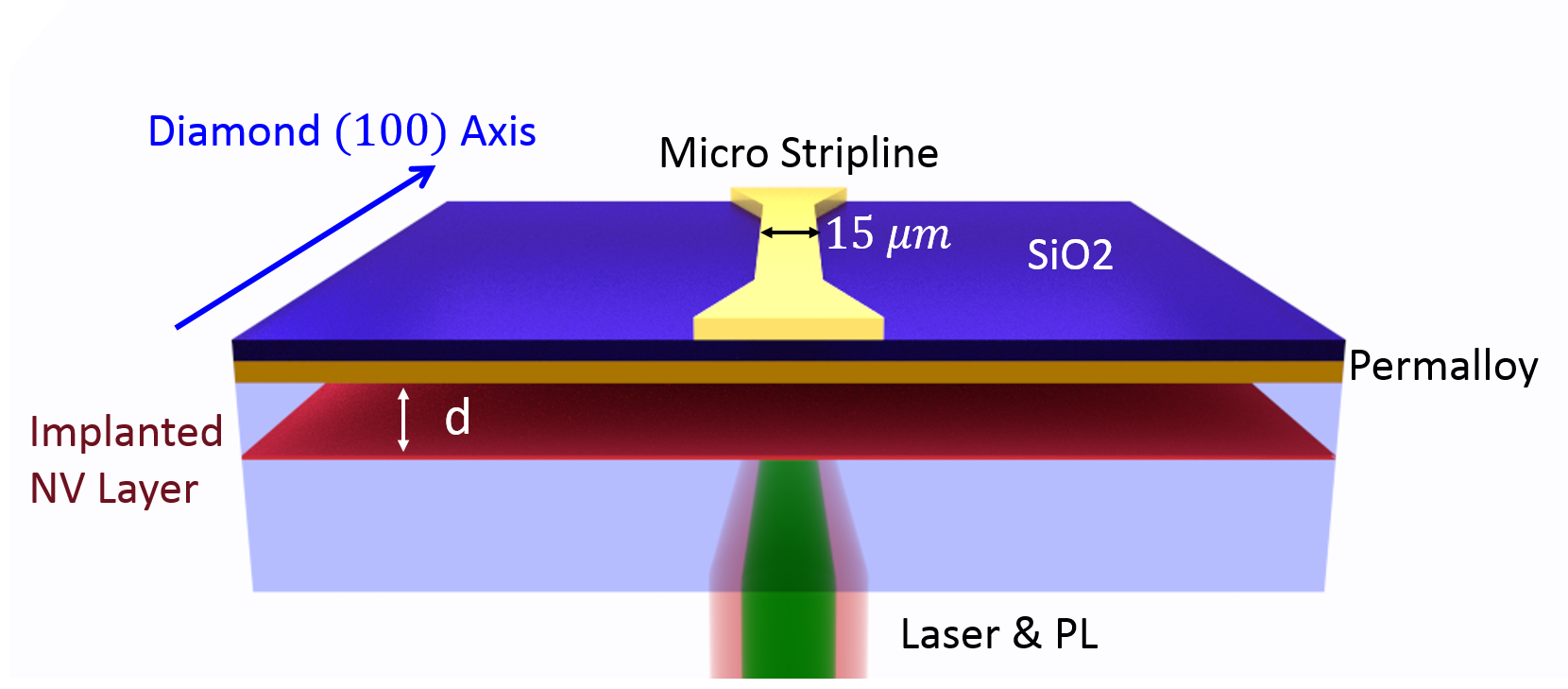}}
\caption{\textbf{A schematic of the experiment.} The sample is a 20 nm Py ferromagnetic film deposited on a single crystal diamond with an implanted layer of NV centers 25 nm - 100 nm from the surface. In order to apply microwave magnetic fields to the sample, a microwire (5 nm Ti/ 300 nm Ag) is patterned on an insulating SiO$_2$ layer. Green laser light is focused through the back of the diamond and the resulting fluorescence changes of the NV centers are monitored. The static field $H_0$ can be applied either in the film plane or along a $\langle$111$\rangle$ axis of NV symmetry.}
\label{fig:schematic}
\end{figure}

We posit that the off-resonant, FMR-induced change in NV center spin-state -- and thus its fluorescence variation -- must result from fluctuating dipolar fields produced by the ferromagnetic excitation. Since the uniform FMR mode of a continuous film cannot produce a fluctuating dipolar field outside its boundaries, we suggest that a spectrum of spatially-inhomogeneous dipole fields from spinwaves are generated during the decay of the uniform mode that relax the NV center spins \citep{schultheiss_direct_2012}. The effect will be largest when the wavelength of a spinwave is comparable to the separation between the surface of the ferromagnet and the NV centers, and when the spinwave frequency matches the NV center spin resonance. Here we present experimental data which supports this idea. 

The experimental results and related analysis are organized as follows: In Section I, we first show an ODFMR signal from a permalloy (Ni$_{80}$Fe$_{20}$) thin film measured using a single crystal diamond which has well-defined NV orientations. In contrast to previous ODFMR measurements using randomly oriented NV centers in nanodiamonds, the single crystal samples containing a thin layer of implanted NV centers allow us to control the NV center-ferromagnet (NV-FM) separation, $d$. We study the variation of the ODFMR signal as a function of $d$. This data suggests that the signal is optimized at a separation matching the wavelength of spinwaves whose frequency matches the NV spin resonance. In Section II, we quantitatively measure the longitudinal spin relaxation lifetime ($T_1$) of the ground-state NV center spin, which is reduced as we drive the FM on resonance. This directly demonstrates that ODFMR arises due to interactions between the ground-state NV center spin and a driven ferromagnetic system. In Section III, we demonstrate the generality of the ODFMR effect by measuring it in three ferromagnets: permalloy (Py), cobalt (Co), and cobalt manganese iron germanium (CMFG) \citep{page_temperature-dependence_2016}, a Heusler alloy of interest for read heads; this helps us probe the mechanism by measuring the variation of the signal with material properties. Finally, in Section IV, we discuss our conclusions and key future experiments.

\section{I: Demonstration of ODFMR in single crystal NV-Py system and dependence of signal on NV-FM separation} 

\begin{figure*}[p]
\center{\includegraphics[width=0.95\linewidth]{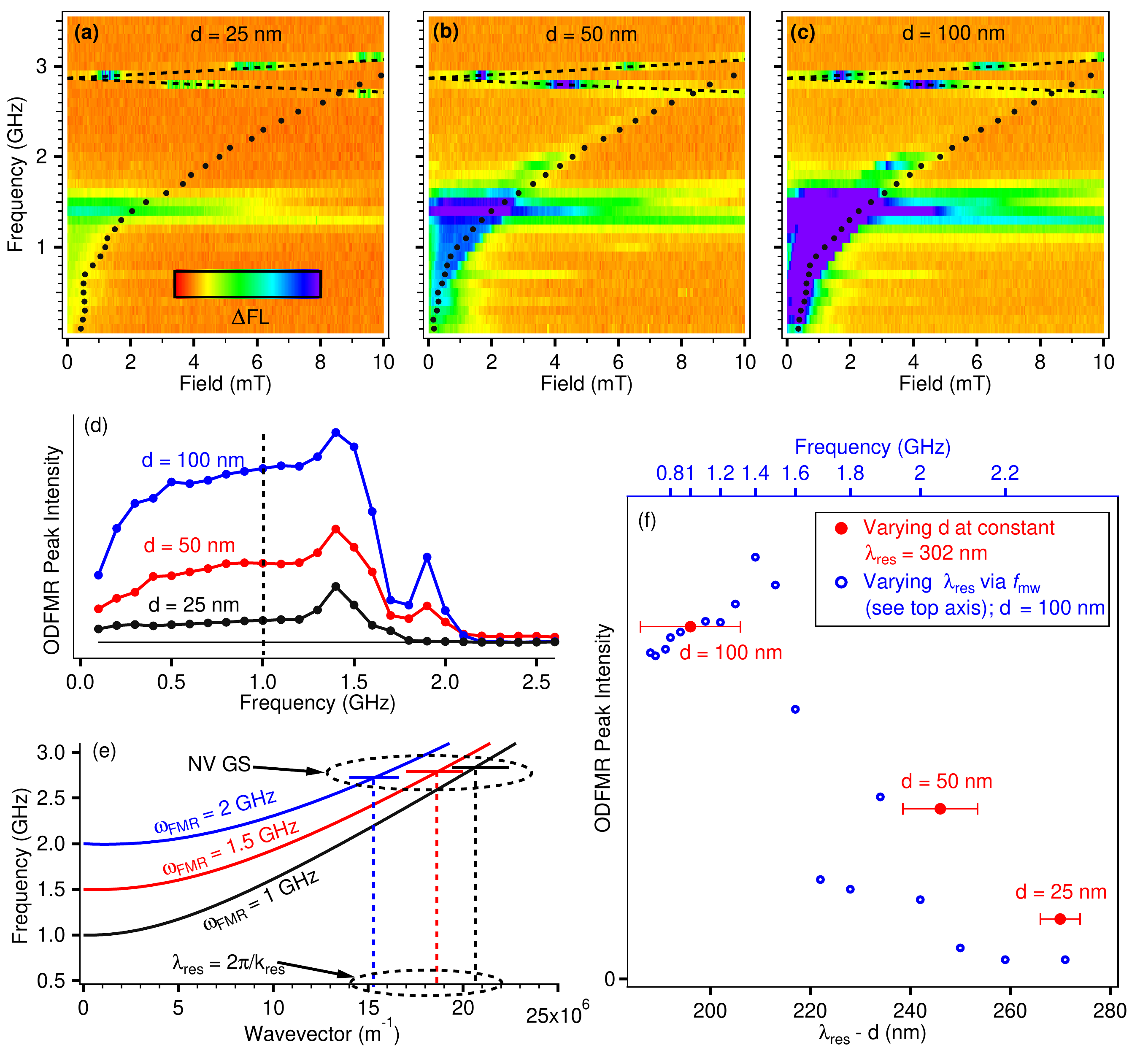}}
\caption {\textbf{ODFMR in Py-single crystal diamond and its dependence on \textit{d}.} CW NV center $\Delta$FL vs $H_0$ and $f_{\text{mw}}$ signals from three single crystal diamonds with NV centers implanted at (a) 25 nm, (b) 50 nm, (c) 100 nm, and capped by a Py film. Black dots indicate the FMR resonance measured inductively. The black dashed lines are a guide to the eye for the ground-state resonances of the NV centers. (d) $\Delta$FL at the ODFMR peak as a function of frequency for three values of $d$. The dashed line indicates the position of the peaks in $\Delta$FL observed at 1 GHz in the three diamond samples, which are re-plotted in panel (f) as red points. Additional ODFMR spectra are shown in the SI. (e) The spinwave dispersion for a 20 nm Py film at the FMR fields for $f_{\text{mw}}$ = 1 GHz, 1.5 GHz, and 2 GHz. The solid horizontal lines indicate the frequency of the NV center resonance at a field corresponding to the respective $f_{\text{mw}}$. The intersection of the spin wave dispersion with the NV center ground state frequency determines the spinwave wavevectors which most efficiently couple to NV centers under FMR excitation. The wavelength of these spin waves, $\lambda_\text{{res}}$, changes as a function of the FMR frequency. See main text for more details. (f) ODFMR peak intensity at the FMR condition of the film versus the normalized distance $\lambda_\text{{res}} - d$, which is varied either by adjusting the FMR frequency (and thus $\lambda_\text{{res}}$, blue), or by varying $d$ in a series of samples (red). The red data show the ODFMR intensity at 1 GHz (see the dashed line in panel (d)), corresponding to $\lambda_\text{{res}}$ = 302 nm, plotted as the implantation depth, $d$ is varied. The blue data are the ODFMR intensity at fixed $d$ = 100 nm (the same data as the blue line in panel (d)), plotted versus the distance $\lambda_\text{{res}} - d$, which is varied by changing the frequency of FMR. $\lambda_\text{{res}}$ is calculated from the dispersion as in panel (e), and the corresponding frequencies are plotted on the top axis. Horizontal error bars are given by the FWHM of ion straggle, extracted from SRIM calculations. All points in blue have the horizontal error bar of the d = 100 nm straggle.}
\label{fig:SingleCrystal}
\end{figure*}

\begin{figure}
\center{\includegraphics[width=0.95\linewidth]{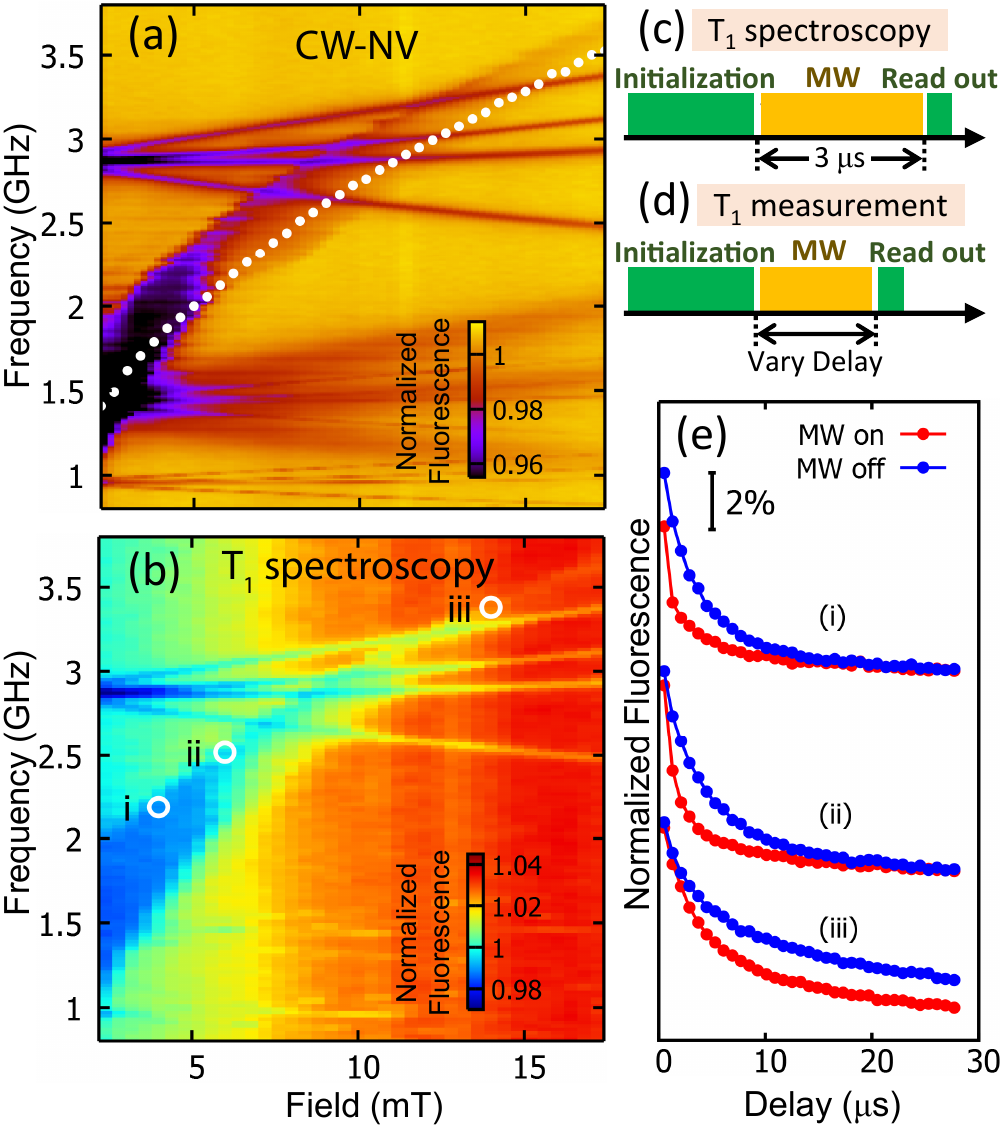}}
\caption{\textbf{Effect of FMR on longitudinal spin-relaxation time.} (a) CW-NV center fluorescence as a function of applied field. The white dots show FMR  measured with reflected microwave power ($S_{11}$). (b) $T_1$ spectroscopy (3 $\mu$s dark time) as a function of applied field. The white open circles show the locations of the measurements in (e). Panel (c) shows pulse sequences for $T_1$ spectroscopy and (d) $T_1$ measurement. (e) $T_1$ measurements with microwave at the FMR frequency on (red) and off (blue).}
\label{fig:Pulsed}
\end{figure}

To unravel the origin of the ODFMR signals, we study its dependence on the controlled separation $d$ between the NV center spins and the surface of a continuous FM film under microwave drive.  If a dipolar spinwave mechanism is relevant, then we expect that spinwaves with a wavevector on the order of $2\pi/d$ will have the largest stray field at the NV center position. Fig. \ref{fig:schematic} presents a schematic of the sample and the geometry used for measurements presented in Figs. \ref{fig:SingleCrystal} and \ref{fig:Pulsed}. The key elements are a single crystal diamond substrate with a layer of NV centers implanted a distance $d$ from the surface. There are also NV centers at a lower density throughout the volume of the diamond, but can be experimentally distinguished from the implanted NV centers, as discussed in the supplementary information (SI). We deposited Py on the diamond surface and then patterned an electrically isolated microstripline antenna to produce a microwave magnetic field, $H_1$. To record ODFMR, we monitor the fluorescence from implanted NV centers as a function of the static magnetic field $H_0$ and the microwave frequency, $f_{\text{mw}}$, of $H_1$. $H_0$ can be oriented in the plane of the Py film or along one of the NV center axes. In addition, we also record the microwave power reflected by the microstripline, $S_{11}$, providing an alternative, spatially averaged measure of FMR. Additional details of each measurement are provided in the SI. 

In Fig. \ref{fig:SingleCrystal}(a)-(c), we show ODFMR measurements of Py films. The continuous wave (CW), normalized fluorescence change ($\Delta$FL) is measured using a lock-in amplifier that is referenced to amplitude modulation of $H_1$. We recorded $\Delta$FL as a function of $H_0$ and $f_{\text{mw}}$ for three single crystal (100) diamond substrates with $d$ = 25 nm, 50~nm, and 100 nm. Black circles overlaid on the 2D color plots show FMR peaks detected by $S_{11}$. The features in $\Delta$FL appearing near $~2.9 $ GHz are the directly-driven ground-state resonances of the NV centers, marked with a guide-to-the-eye. With $H_0$ aligned in the plane of the sample along the $\langle$100$\rangle$ axes, the four orientations of NV center $\langle$111$\rangle$ symmetry axes are degenerate in this measurement set-up. The feature around 1.2 GHz-1.6 GHz and extending over the measured field range is due to the NV center excited-state spin resonances, which are broader than ground-state, and thus overlap \citep{fuchs_excited-state_2008}. The ODFMR peaks due to the Py FMR appear most clearly in the low-field, low-frequency regime.  Fig. \ref{fig:SingleCrystal}(d) also shows that as $d$ increases, the intensity of the ODFMR signal also increases. Furthermore, the ODFMR signal is consistent with the direct measurement of $S_{11}$, shown along the black points. However, as a subtle but intriguing point, the ODFMR peaks are centered at lower fields than the $S_{11}$ detected FMR peaks.  This may merely be related to differences in the local and global FMR, since we probe NV centers at the narrowest point of the stripline, and thus they experience a larger $H_1$ than the majority of the FM. Alternatively, this may provide valuable clues as to the spectral efficiency of spinwave generation that produce ODFMR.

As discussed above, we hypothesize that  ODFMR results from fluctuating dipolar fields due to spinwaves that decay from the driven, uniform FMR mode. Here we discuss the spinwave dispersion in relation to FMR frequency. Fig. \ref{fig:SingleCrystal}(e) shows representative dispersions for the spin waves in Py for uniform mode FMR frequencies of 1 GHz, 1.5 GHz, and 2 GHz. The uniform mode can decay into spinwaves with frequencies and wavevectors that are allowed by this dispersion \citep{lukaszew_handbook_2015}. We use this dispersion to estimate the wavelength (see the SI for details), $\lambda_\text{{res}}$, of the spinwave whose frequency is resonant with the $\ket{0}$ $\rightarrow$ $\ket{-1}$ NV center spin transition for each frequency of the uniform mode FMR. 

To summarize the importance of matching $\lambda_\text{{res}}$ to $d$, in Fig. \ref{fig:SingleCrystal}(f) we present the measured ODFMR peak intensity from Fig. \ref{fig:SingleCrystal}(d), re-plotted as a function of $\lambda_\text{{res}} - d$. This difference can be varied either through $\lambda_\text{{res}}$, by changing the uniform FMR resonance frequency (e.g. by changing the magnetic field), or through $d$, by changing the NV center implantation depth. The blue points show the intensity at $d$ = 100 nm, but plotted versus $\lambda_\text{{res}} - d$ using the spinwave dispersion to convert the horizontal coordinate. The correlation between $\lambda_\text{{res}}$ and frequency is shown in the top axis. The red points are measured using a 1 GHz microwave drive, so $\lambda_\text{{res}}$ = 302 nm for each of the three samples ($d$ = 25 nm, 50 nm, 100 nm). The variation of the ODFMR signal obtained by these two methods shows the qualitative trend that we expect from our hypothesis: the signal decreases as the difference between $\lambda_\text{{res}}$ and $d$ grows, regardless of the method by which $\lambda_\text{{res}} - d$ is changed. Note that we have removed the data in the extremely low-field regime in which the film is in a multi-domain state. Additionally, we point out other effects that appear in the ODFMR signal. For example, the peak around 210 nm arises from the coincidence of the NV center excited-state resonance and the FMR frequency, while the dip around 230 nm is an artifact due to reduced $H_1$ arising from a standing wave resonance in our microwave circuit.

\section{II: Effect of FMR on NV center longitudinal spin-relaxation time and T$_1$ spectroscopy}

Fluctuating dipolar fields of spinwaves will enhance the NV spin relaxation rate \citep{van_der_sar_nanometre-scale_2015}. Here we present $T_1$
measurements that demonstrate a reduction of the NV center $T_1$ in response to excitation of FMR.

We expect that the fluorescence change in ODFMR can be interpreted as a change in the NV center spin populations as we drive FMR. By measuring the longitudinal spin-relaxation lifetime of the NV center ground spin state with and without FMR, we can quantify how driving FMR modifies the NV spin populations. Using the $d = 50$ nm sample described above, we first characterize CW ODFMR in a different measurement set-up that is configured to measure the lifetimes of the NV centers, shown in Fig. \ref{fig:Pulsed}. In this setup, $H_0$ is applied along one of the $\langle$111$\rangle$ NV center axes and the fluorescence is monitored using a single photon counter. For this field orientation the four NV center ground state branches are visible, in contrast to the two for in-plane field (Fig. \ref{fig:SingleCrystal}). Next we perform a $T_1$ spectroscopy measurement using the pulse sequence shown in Fig. \ref{fig:Pulsed}(c). The NV centers are first initialized in the $\ket{0}$ spin state with a laser pulse, and following a fixed delay time of 3 $\mu$s, we pulse the laser again for fluorescence readout. During the 3 $\mu$s dark time, a microwave pulse is applied with the same power as in Fig. \ref{fig:Pulsed}(a). We sweep the microwave frequency at fixed field, then repeat the $T_1$ spectroscopy measurement as a function of applied field, as shown in Fig. \ref{fig:Pulsed}(b). 

$T_1$ spectroscopy is an effective approach to quickly find the conditions in which $T_1$ changes. Two main features are displayed in Fig. \ref{fig:Pulsed}(b). First, when the FMR occurs, we observe a significant reduction in the NV center $T_1$. As in Fig. \ref{fig:Pulsed}(a), the FMR signal is pronounced in the low field/frequency region.  At high field/frequency where the FMR frequency exceeds the NV center resonance frequency, the FMR signal gradually diminishes. The second feature in Fig. \ref{fig:Pulsed}(b) is that the background of $T_1$ spectroscopy increases with increasing field. The field dependent $T_1$ spectroscopy background is consistent with a separate measurement of $T_1$ vs. field without the microwave drive. The effect has been previously studied in an undriven Py/diamond system where it was attributed to relaxation by thermal fluctuations of the Py magnetization that produces wide spectrum spin noise \citep{van_der_sar_nanometre-scale_2015}. These measurements directly demonstrate that the ODFMR signal arises due to fluctuating dipolar fields generated by a driven ferromagnetic system that relax the ground-state NV center spins, without requiring participation of the NV center excited state.

\begin{figure}[t]
\center{\includegraphics[width=0.85\linewidth]{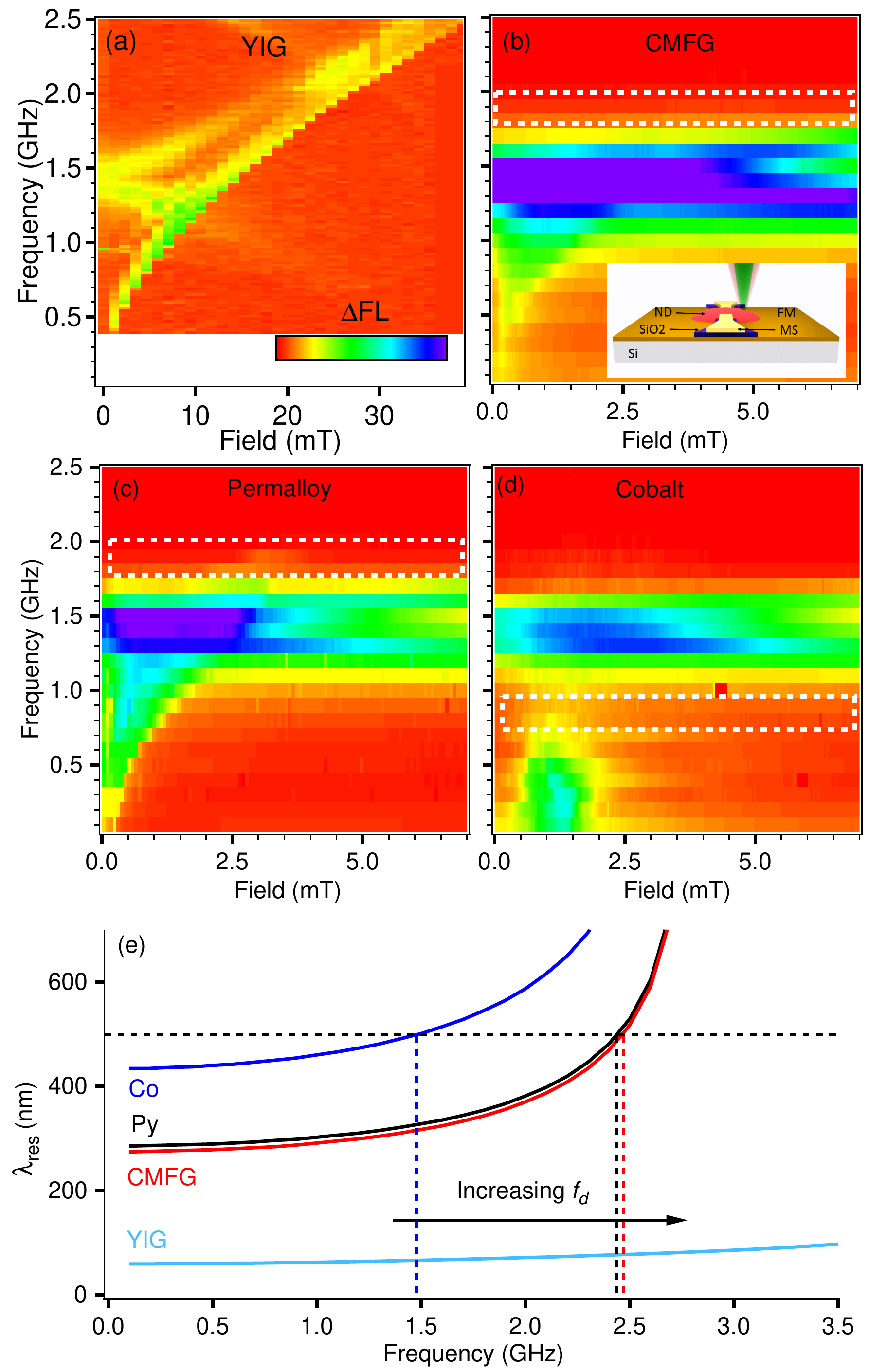}}
\caption{\textbf{The effect of material properties on ODFMR.} $\Delta$FL as a function of field and frequency of NV centers in a nanodiamond (ND) film on top of various ferromagnetic films. Inset: the device schematic for the samples measured using nanodiamonds; see SI for details. In (a)-(d), four different ferromagnetic films are shown, YIG, CMFG, Py, and Co respectively. The dashed white boxes are a guide for the frequency, $f_\text{d}$, for which the ODFMR signal decays. The data for YIG are from our previous publication \citep{wolfe_off-resonant_2014}, and no decay frequency is seen in the measured frequency range. (e) The optimal distance for the NV centers determined from matching of the spinwave modes to the NV resonance condition as a function of frequency for Co, Py, CMFG, and YIG. This is calculated as in Fig. \ref{fig:SingleCrystal} (e).  Changes in the saturation magnetization and the exchange stiffness affect the spin wave dispersion, which in turns affects the spin wave wavelength resonant with the NV ground state. The horizontal black dashed line indicates the largest NV-FM separation, $d$, encountered in our nanodiamond films. For $\lambda_\text{{res}}$
 larger than the maximum $d$, the signal intensity will decay; the corresponding calculated frequency maximum is shown by the vertical dashed lines. These data are suggestive of qualitative agreement between the decay frequency and the frequency at which  $\lambda_\text{{res}}$
 surpasses $d$.}
\label{fig:Comp}
\end{figure}

To better quantify the $T_1$ change during FMR, we also directly measure $T_1$ with and without a microwave drive. We choose three points in Fig. \ref{fig:Pulsed}(b) for $T_1$ measurement, as shown in Fig. \ref{fig:Pulsed} (e). When the FMR condition is met, a large reduction in $T_1$ is observed, in agreement with our $T_1$ spectroscopy results. For example, at 40 G (curve (i) in Fig. \ref{fig:Pulsed}(e)) the measured $T_1 = 3.8 \pm 0.1$ $\mu$s with no microwave applied. However, when a microwave of 2.19 GHz is applied to satisfy the FMR condition, $T_1=2.1\pm0.2$ $\mu$s, which has reduced nearly by a factor of 2.  Despite the fact that the FMR frequency is not resonant with the NV center spin resonances, the $T_1$ reduction during FMR suggests an incoherent coupling mechanism between NV centers and the nearby driven FM. This hypothesis pertains to the $\Delta$FL as a result of driving the uniform mode. Changes in the fluorescence of NV centers as a result of thermal equilibrium spin waves in a nearby Py disk in the undriven case have already been investigated \citep{van_der_sar_nanometre-scale_2015}.

\section{III: Demonstration of ODFMR in different ferromagnets and dependence of the signal on material properties}

The spinwave dispersion is sensitive to saturation magnetization, $M_s$, and exchange stiffness, $A$, making this phenomenon sensitive to material-specific details of the FM. Here we describe differences in ODFMR as a function of material properties, and their relationship to our
proposed mechanism.

The sample structure is shown in the inset of Fig. \ref{fig:Comp} and additional details can be found in the SI. $\Delta$FL collected from a nanodiamond film was measured using a lock-in amplifier and is presented in Fig. \ref{fig:Comp} as a function of in-plane $H_0$ and $f_{\text{mw}}$ for four samples: a 30 nm YIG film on GGG from \citep{wolfe_off-resonant_2014}, a 20 nm Py film on Si, a 5 nm CMFG film on glass, and a 20 nm Co film on Si. The white dashed boxes highlight the frequencies, $f_\text{d}$, above which the ODFMR signal decays for each material.  This frequency is lowest for the Co ODFMR signal, with $f_\text{d}$ $\sim 0.7$ GHz, and it is larger for Py and CMFG which exhibit $f_\text{d} \sim 1.8$ GHz, while the YIG ODFMR signal does not strongly decay in this frequency range. Note that the field scale in panel (b)-(d) is smaller than in (a). In Fig. \ref{fig:Comp}(e) the calculated $\lambda_\text{{res}}$
 is plotted as a function of the FMR frequency for each material. The values of $M_s$ and $A$ in the various materials affect their spinwave dispersions, and hence the wavelengths of the spinwaves that are resonant with the NV center ground state spin transition. The horizontal black dashed lines indicate the value of $d$ that is expected based on the thickness of the nanodiamond films. When $\lambda_\text{{res}}$ becomes larger than this limiting value of $d$, we expect that the signal intensity will decay. The estimated decay frequencies are shown by the vertical dashed lines. We find that $f_\text{d}$ (Co) $<$ $f_\text{d}$ (Py) $\approx$ (CMFG) $<$ $f_\text{d}$ (YIG), which matches the experimental trend for the ODFMR cut-off frequency in each material. 

\section{IV: Conclusions and outlook}

While our hypothesis qualitatively explains the results presented here and in the previous work \citep{wolfe_off-resonant_2014}, a more quantitative theory of the NV-FM coupling is still needed. Noise arising from the incoherent bath of spinwaves reduces the NV center spin lifetime as shown in Fig. \ref{fig:Pulsed}. Thus measuring the spin lifetime $T_1$ of the NV centers on and off FMR as a function of $d$ is an important future step. The change of $T_1 $ is related to the strength of the fluctuating dipolar fields generated by these spinwaves, which can be calculated theoretically and be quantitatively compared with our measurements. A more detailed theory will also include the spectral density of spinwaves generated by the decay of the uniform mode. An important aspect of this work is the possible avenues it highlights for experimentally characterizing damping processes in ferromagnets, such as magnon-magnon interactions, which are likely responsible for generating higher energy spinwaves from the uniform mode. Understanding this phenomenon could lead to a new method for deducing the rate at which spinwaves are generated by magnetization dynamics of ferromagnets, with a sensitivity that can be selected by controlling the separation between the NV centers and FM surface to match the desired properties of spinwave to be measured.  Extending this idea to scanned-probe sensing in dynamical magnetic systems will offer a unique modality for nanoscale magnetic imaging.

In conclusion, we have demonstrated the optical detection of ferromagnetic resonance in a variety of ferromagnets using NV centers. We hypothesize that spinwaves generated by decay of the uniform mode that match the NV center resonance are responsible for altering the NV center spin lifetime, thus generating an ODFMR signal. Although our data qualitatively support this hypothesis, it requires further measurements of the NV spin lifetime and a detailed theoretical analysis. 

We thank Dr. Sergei Manuilov for helpful discussion. Funding for this research at The Ohio State University was provided by the ARO through award number W911NF-12-1-0587 and the Center for Emergent Materials at the Ohio State University, an NSF MRSEC through award Number DMR-1420451. 
We acknowledge use of Ohio State Nanosystems Laboratory shared facilities. Research at Cornell is supported in part by the AFOSR (grant \# FA9550-14-1-0243). We acknowledge use of the shared facilities of the Cornell Center for Materials Research under grant DMR-1120296.

\bibliography{OpticallydetectedFMRCitations}

\end{document}


\title{Supplementary Material: Optically Detected Ferromagnetic Resonance in Metallic Ferromagnets via Nitrogen Vacancy Centers in Diamond}

\author{M. R. Page*}
\affiliation{Department of Physics, The Ohio State University, Columbus, Ohio 43210, USA}

\author{F. Guo*}
\affiliation{School of Applied and Engineering Physics, Cornell University, Ithaca NY 14850}

\let\thefootnote\relax\footnote{* M. R. Page and F. Guo contributed equally to this work.}

\author{C. M. Purser}
\affiliation{Department of Physics, The Ohio State University, Columbus, Ohio 43210, USA}

\author{J. G. Schulze}
\affiliation{Department of Physics, The Ohio State University, Columbus, Ohio 43210, USA}

\author{T. M. Nakatani}
\affiliation{San Jose Research Center, HGST, a Western Digital company, San Jose, California 95135, USA}

\author{C. S. Wolfe}
\affiliation{Department of Physics, The Ohio State University, Columbus, Ohio 43210, USA}

\author{J. R. Childress}
\affiliation{San Jose Research Center, HGST, a Western Digital company, San Jose, California 95135, USA}

\author{P. C. Hammel}\email{hammel@physics.osu.edu}
\affiliation{Department of Physics, The Ohio State University, Columbus, Ohio 43210, USA}

\author{G. D. Fuchs}\email{gdf9@cornell.edu}
\affiliation{School of Applied and Engineering Physics, Cornell University, Ithaca NY 14850}

\author{V. P. Bhallamudi}
\affiliation{Department of Physics, The Ohio State University, Columbus, Ohio 43210, USA}

\date{\today}


\maketitle

\section{Creation of single crystal substrates with thin implanted layer of NV centers}

The single crystal diamond samples are 3 mm x 3 mm x 0.3 mm optical grade, CVD diamonds from Element 6 with [N] $<$ 1ppm, [B]$<$ 0.05, (100) oriented faces, and $<$100$>$ oriented edges. The diamonds were implanted with a thin layer of NV centers by bombardment with $^{15}$N at 18 keV, 40 keV, and 87 keV to achieve mean depths of 25 $\pm$ 8 nm, 50 $\pm$ 13 nm, and 100 $\pm$ 20 nm according to SRIM simulations. The 25 nm-deep sample was implanted with fluence 3.53 x10$^{13}$ atoms/cm$^2$ to achieve a peak nitrogen density of 100 ppm. The 50 and 100 nm sample doses were 1.07x10$^{14}$ atoms/cm$^2$ and 1.54x10$^{14}$ atoms/cm$^2$ to achieve peak nitrogen densities of 200 ppm. In order to convert implanted nitrogen centers into NV- centers, the samples were annealed at 915 C for 2 hours in an atmosphere of approximately 350 mTorr Ar and 50 mTorr H$_2$ gas.

\begin{figure}
\center{\includegraphics[width=0.9\linewidth]{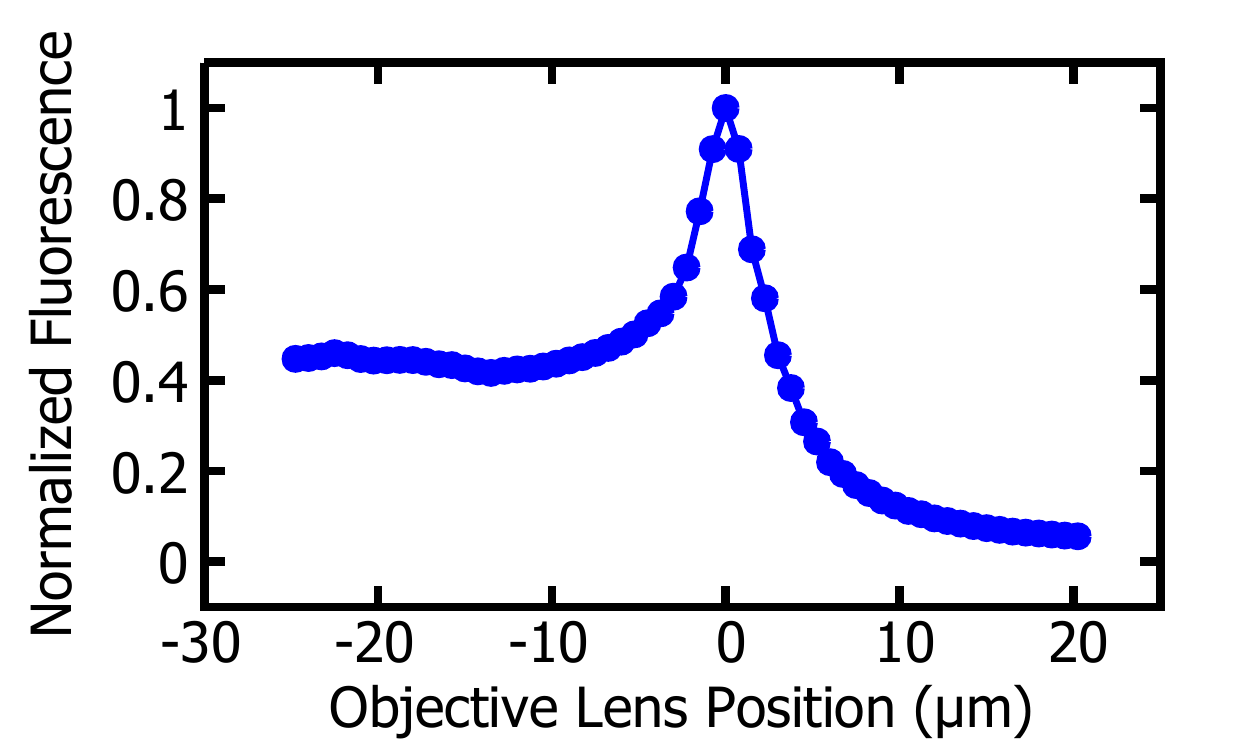}}
\caption{Fluorescence measured as a function of objective lens position.}
\label{fig:BNV}
\end{figure}

We note that there are native NV centers throughout the entire diamond crystal prior to the implantation. The ion implantation introduced an additional layer of NV centers. The implanted and native NV centers can be distinguished by measuring the depth dependent fluorescence, as shown in Fig.~\ref{fig:BNV}. The normalized fluorescence intensity of NV centers is recorded as a function of the objective lens position. When the green pump laser is focused well inside the bulk of the diamond crystal, corresponding to a negative position of the objective lens, we detect a nearly constant fluorescence independent of the objective's position due to the uniform distribution of native NV centers inside the diamond crystal. While the focus of the green light approaches the diamond surface where the additional layer of NV centers is implanted, a fluorescence peak is observed. As the objective lens continues to move towards positive direction, the fluorescence level quickly decays to zero as the pump laser is focused outside of the diamond. We point out that the width of the fluorescence peak is limited by the depth of focus, and it is much wider than the true distribution of the implanted NV centers. For the data of Fig.~3 in the main text, the position of the objective is fixed at 0. Therefore the majority of the fluorescence signal comes from the implanted layer of the NV centers.

\section{Ferromagnetic film deposition and fabrication of microwave antenna}

\begin{figure}
\center{\includegraphics[width=0.9\linewidth]{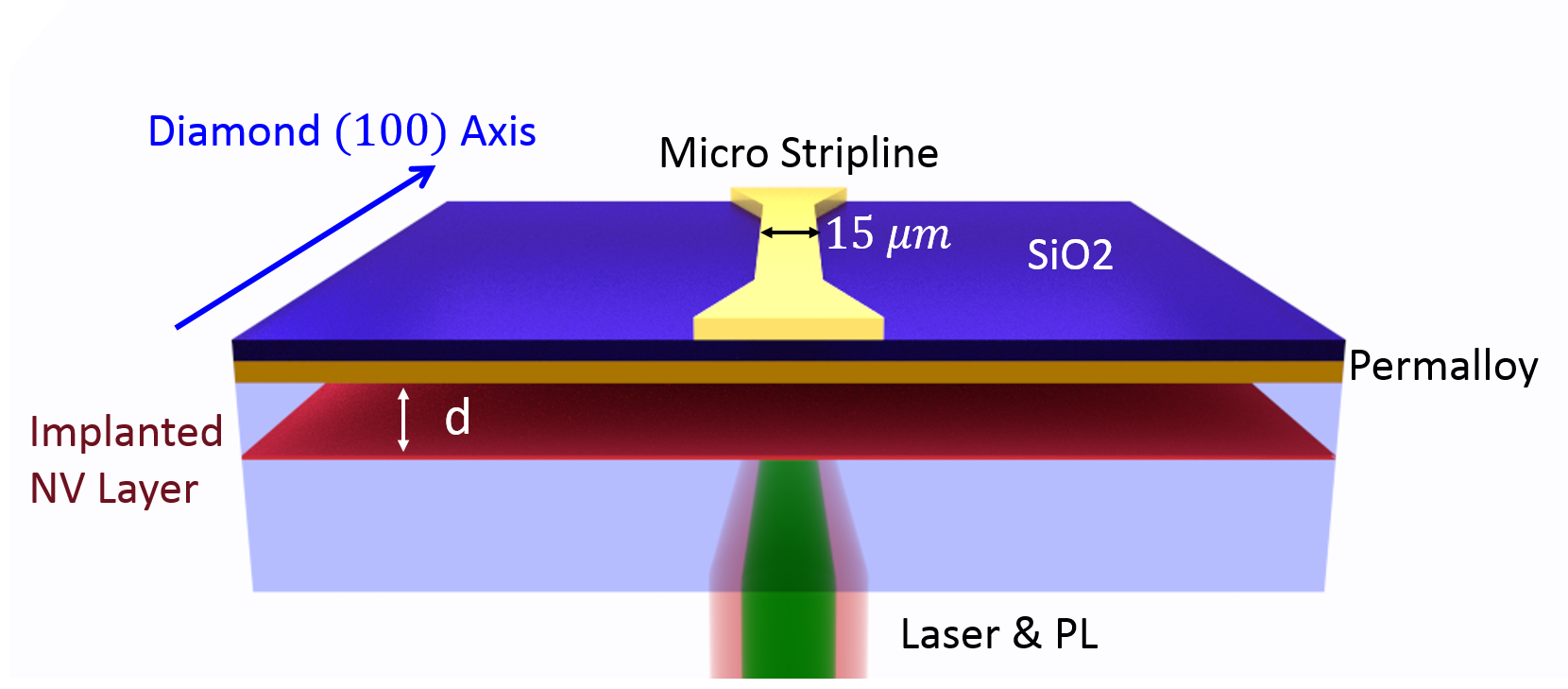}}
\caption{Experimental schematic: The sample is a 20 nm Py ferromagnetic film deposited on a single crystal diamond with an implanted layer of NV centers 25-100 nm from the surface. In order to apply microwave magnetic fields to the sample, a microwire (5 nm Ti/ 300 nm Ag) is grown on an insulating SiO$_2$ layer. Green laser light is focused through the back of the diamond and the changes in the resulting fluorescence of the NV centers is monitored as a function of applied external magnetic field and microwave excitation frequency. The applied field $H_0$ can be applied either in the film plane or along one axis of the NV center.}
\label{fig:schematic}
\end{figure}

\begin{figure}
\center{\includegraphics[width=0.6\linewidth]{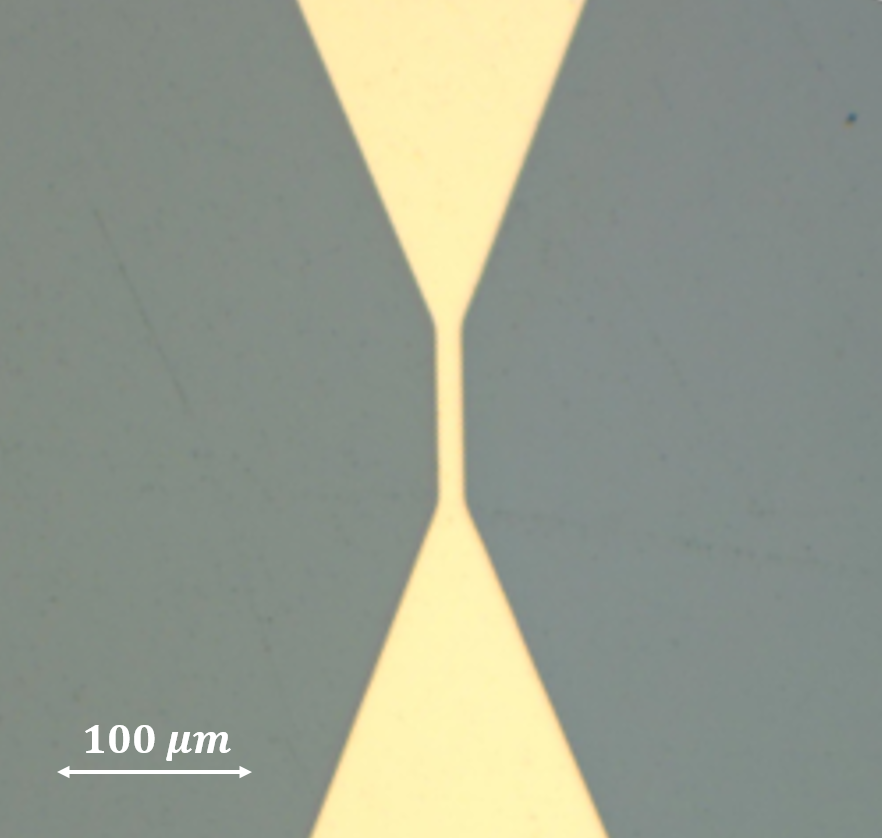}}
\caption{Optical image of the $d$ = 50 nm single crystal diamond sample with Py and miccrostripline. The Py film and SiO$_2$ layer cover the top surface of the diamond. The patterned microstripline can be seen in gold.}
\label{fig:OptIm}
\end{figure}

A schematic of the single crystal diamond samples is shown Fig. \ref{fig:schematic}, and an optical image of the diamond with $d$ = 50 nm is shown in Fig. \ref{fig:OptIm}. For these samples, a 20 nm thin layer of Py is first deposited by electron beam evaporation on the surface of the diamond near the implanted NV center layer. After using RF sputtering for the deposition of 150~nm of an insulating SiO$_2$ layer on top of the Py, a microstrip is lithographically patterned on the sample to apply a microwave frequency magnetic field. The microstripline consists of 5 nm Ti, 285 nm Ag, and 10 nm of Au all deposited by electron beam evaporation. For the nanodiamond samples, the metallic ferromagnetic films are deposited on a suitable substrate. Py and Co are deposited by electron beam evaporation to a thickness of 20 nm on Si and CMFG is sputtered to a thickness of 5 nm on glass. The 150 nm SiO$_2$ layer and the 300 nm antenna are patterned on top of the ferromagnet. A film of nanodiamonds of approximately 500 nm in thickness is drop-cast on the sample above the ferromagnetic film.

\section{Continuous wave lock-in based measurements}

\begin{figure*}
\center{\includegraphics[width=0.95\linewidth]{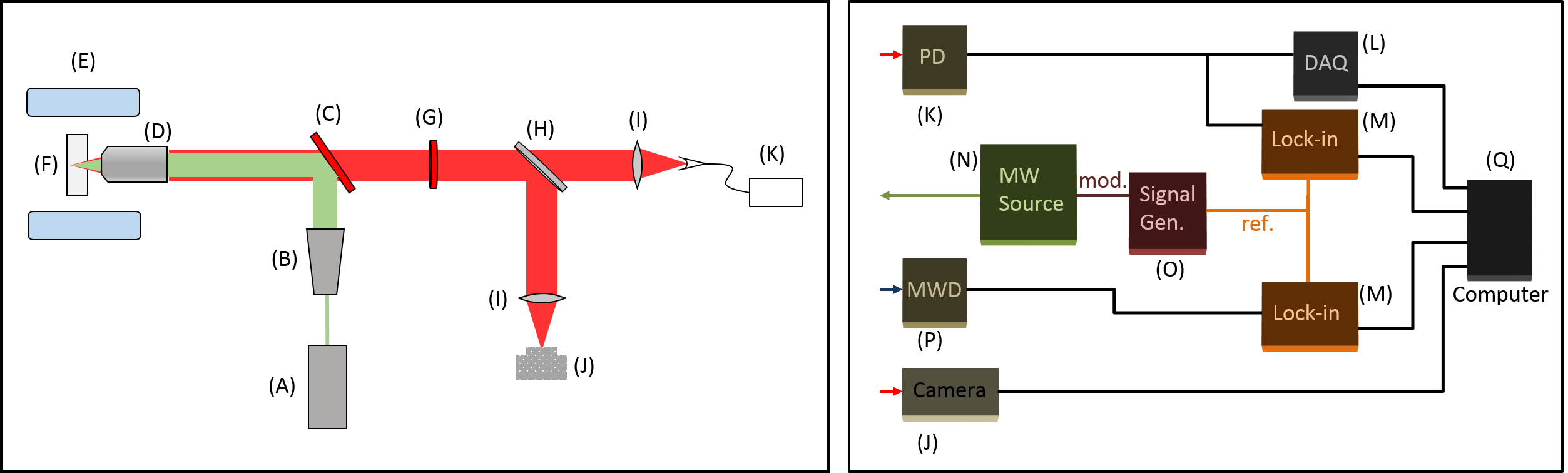}}
\caption{Block diagram for the optical components and instrumentation used in the CW lock-in measurements. Figure adapted from \citep{wolfe_novel_2015}. See Table \ref{tab:BD} for description of each component.}
\label{fig:BD}
\end{figure*}

For the CW measurement of ODFMR in the single crystal and nanodiamond samples from Figs. 2 and 4 in the main text, fluorescence is excited in the NV centers using a 520 nm laser from Coherent model OBIS 520 LX, focused down $<2\,\mu$m spot using a Nikon CFI Plan Fluor 40x objective with 0.75 N.A.  with 35 mW of incident power, and measured by a Laser Components A-CUBE S1500-10 avalanche photodiode. The NV center fluorescence is recorded at a spot under the microstrip for the single crystal samples, and next to the microstripline for the nanodiamonds, as a function of the strength of the static magnetic field $H_0$, applied in-plane and parallel to the antenna, and the frequency, $f_{\text{mw}}$, of the applied microwave field $H_1$. A shorted microstrip geometry is used to produce a microwave field with a Wiltron 68147B signal generator. For the single crystal diamonds, the microstripline is shorted and connected to a 50 $\Omega$ transmission line by indium pressed wires, for the nanodiamonds, the wires are wirebonded. We measure a lock-in signal using a Signal Recovery 7256 for both the fluorescence and the reflected microwave power ($S_{11}$) by modulating the amplitude of $H_1$ at $\sim 1$ kHz. $S_{11}$ allows simultaneous inductive measurement of the FMR using a Krytar 303BK diode, albeit spatially averaged. The raw voltage from the photodiode lock-in is normalized by dividing the voltage at the NV center ground state resonance peak at 0 field at 2.9 GHz. A block diagram of the various optical and measurement hardware components used is shown in Fig. \ref{fig:BD} with description of each component in Table \ref{tab:BD}.

\begin{table}
\caption{Optical components and measurement hardware}
\begin{tabular}{l l c}

\textbf{Letter} & \textbf{Description} \\ 
A & pump laser  \\ 
B & beam expander  \\ 
C & dichroic mirror  \\ 
D & objective lens  \\ 
E & magnet & \\ 
F & sample  \\ 
G & color filters  \\ 
H & 50$/$50 mirror  \\ 
I & lens   \\ 
J & camera  \\ 
K & photodiode \\ 
L & DAQ  \\ 
M & lock-in  \\ 
N & MW source  \\ 
O & signal generator  \\ 
P & MW diode  \\ 
Q & computer  \\ 
\end{tabular}
\label{tab:BD}
\end{table}

\section{NV center Lifetime measurements}

\begin{figure*}
\center{\includegraphics[width=0.95\linewidth]{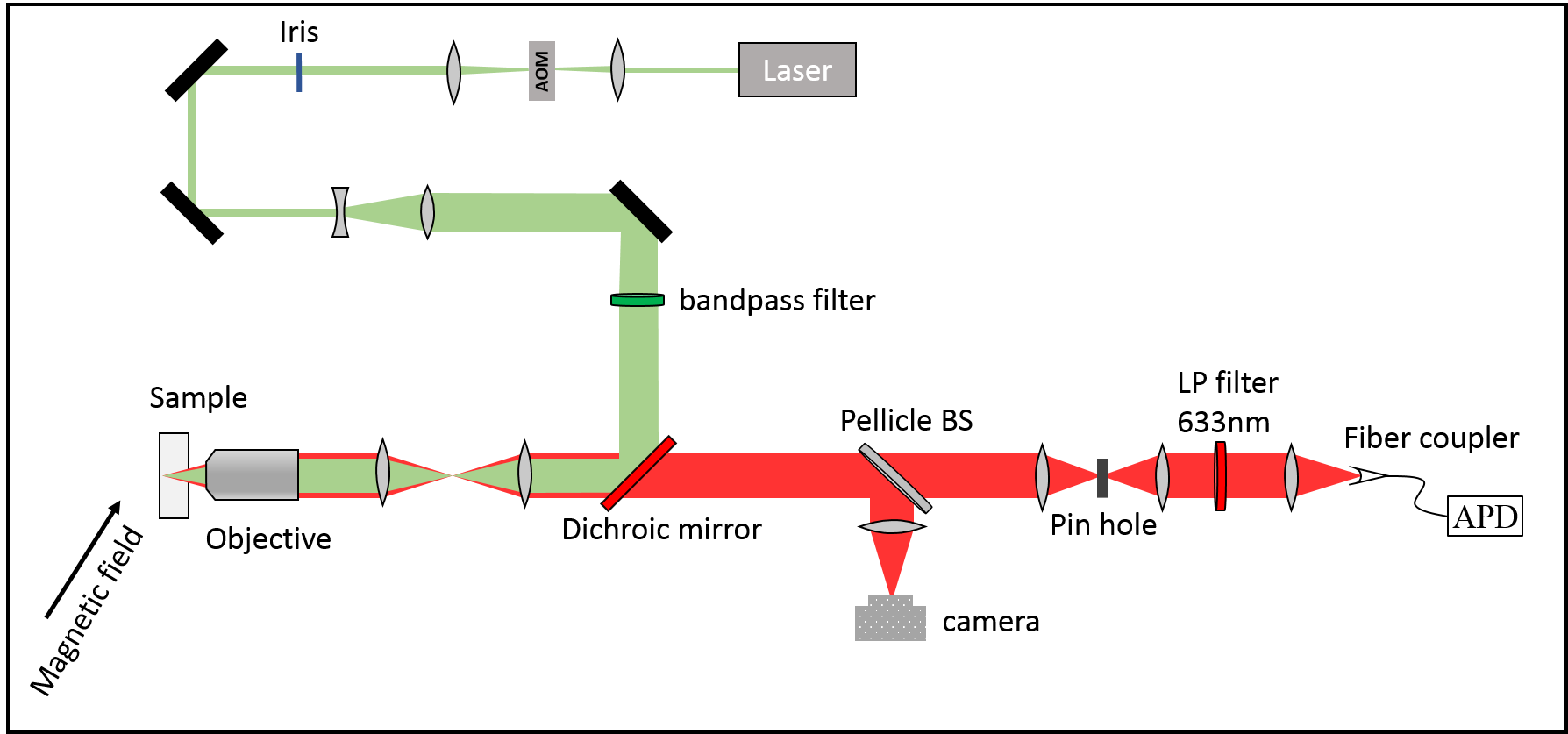}}
\caption{Block diagram for the optical components used in the lifetime measurements.}
\label{fig:PBD}
\end{figure*}

The measurements in Fig. 3 from the main text are performed using a home-built confocal microscopy, as illustrated in Fig. \ref{fig:PBD}. We use a 532~nm green laser for initialization and read out. An ISOMET 1250C acousto-optic modulator (AOM) is used as an optical shutter. Both excitation laser and NV center fluorescence are focused through an Olympus LCPLFLN 50x objective. The fluorescence emitted from NV centers is detected with an Excelitas SPCM-AQR-H-13-FC avalanche photodiode (APD) and a home made RF switch box. The RF magnetic microwave is generated by a SRS SG384 signal generator using IQ modulation. We use a permanent magnet to produce the magnetic field. The magnet is mounted on a transitional stage and a home-built goniometer to control the strength and orientation of the magnetic field.

\section{ODFMR Spectra}

\begin{figure*}
\center{\includegraphics[width=0.9\linewidth]{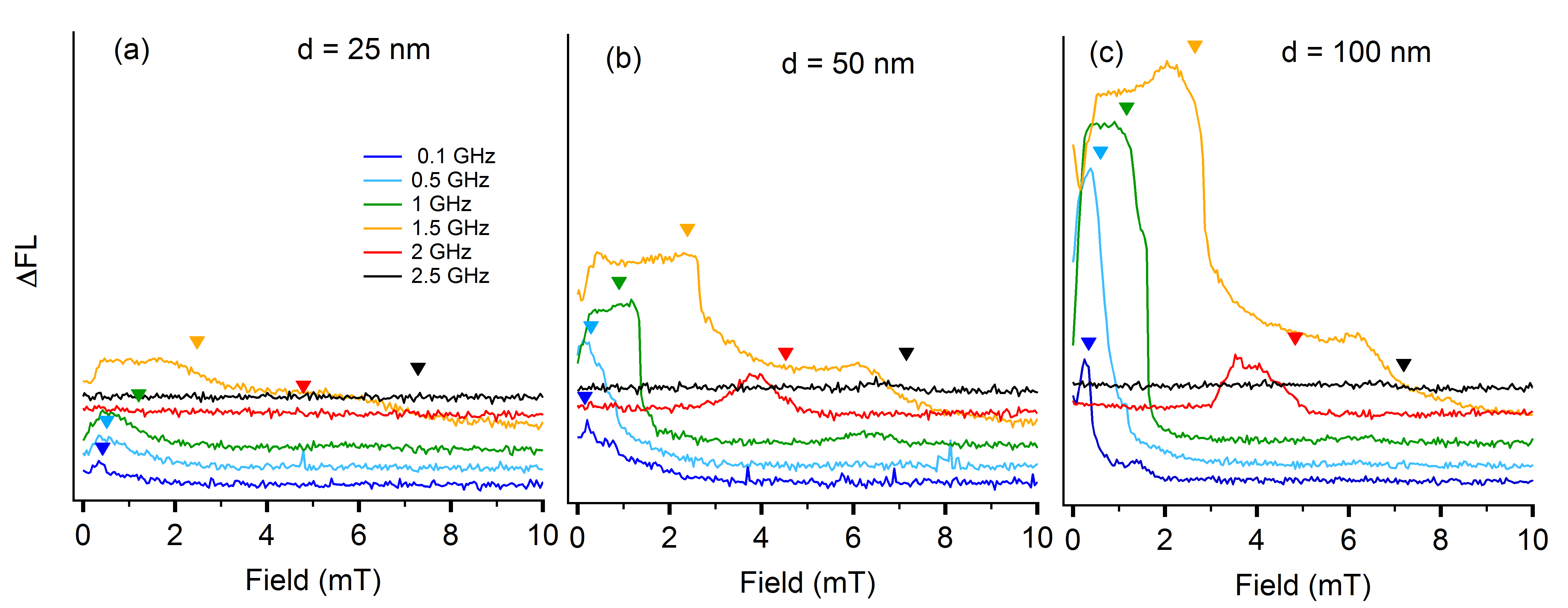}}
\caption {$\Delta$FL as a function of field for several frequencies for the diamond with $d$ = 25 nm (a), $d$ = 50 nm (b), and $d$ = 100 nm (c). The inverted triangles indicate the position of the ODFMR peak as determined from Lorentzian fits to the S$_{11}$ detected FMR peak.}
\label{fig:spectra}
\end{figure*}

Sample ODFMR spectra are shown in Fig. \ref{fig:spectra}. These spectra are line cuts through the 2D plots presented in the main text. In addition, the ODFMR peak, marked with triangles, is recorded as a function of frequency for each of the implantation depths for the analysis in the main text. The position of the triangle is determined from Lorentzian fits to the S$_{11}$ detected FMR peak.  

\section{Calculation of the optimal wavelength}

As discussed in the main text, we use the spinwave  dispersion to theoretically determine the wavelength, $\lambda_\text{{res}}$, of the spinwave whose frequency is resonant with the NV center for each frequency of the uniform mode FMR. This is calculated as follows. The dispersion of spin waves for the uniform mode is given by \citep{lukaszew_handbook_2015} :

\begin{equation}
\left(\frac{\omega(k_\parallel)}{\gamma}\right)^2 = (\frac{\omega_{FMR}}{\gamma})^2 - H_d\pi M_st_Fk_\parallel+H_eD_sk_\parallel^2
\label{eqn:1}
\end{equation}

\begin{equation}
(\frac{\omega_{FMR}}{\gamma})^2 = H(H+4\pi M_{eff})
\label{eqn:2}
\end{equation}

\begin{equation}
H_d=2(H-(H+4 \pi M_{eff})\sin^2\phi_k)
\end{equation}

\begin{equation}
H_e=2H+4 \pi M_{eff}
\end{equation}

\noindent where $\omega(k_\parallel)$ is the frequency of the spin wave, $\gamma$ is the gyromagnetic ratio, $\omega_{FMR}$ is the frequency of the uniform mode, $H$ is the external magnetic field, $M_s$ is the saturation magnetization, $t_F$ is the film thickness, $M_{\text{eff}}$ is the effective magnetization, $k_\parallel$ is the spinwave wavevector, $D_s = 2A / M_s$ where $A$ is the exchange stiffness and $\phi_k$ is the angle between the spinwave wavevector and magnetization.

\noindent For each uniform mode FMR frequency, the Kittel equation shown in Eqn. \ref{eqn:2} above is used to determine the resonant field. Using this resonant field, the NV center ground state resonance condition is determined by finding the eigenvalues of the NV center Hamiltonian \citep{schirhagl_nitrogen-vacancy_2014} below. 

\begin{equation}
\cal{H_{NV}}  = \mathnormal{\gamma H \cdot S +\cal{D}\mathnormal{(S_z \cdot S_z)}}
\end{equation}

\noindent where $\cal{D}$ is the 0 field crystal splitting equal to approximately 2.87 GHz at room temperature, $S$ is the spin 1 matrix with components $S_x$, $S_y$ and $S_z$, and $\vec{H}$ is the vector external magnetic field defined as

\begin{equation}
\vec{H} = H_0(\cos(\theta)\sin(\phi)\hat i, \sin(\theta)\sin(\phi)\hat j, \cos(\phi)\hat k) 
\end{equation}

\noindent The constants $\theta$ and $\phi$ are the relative angles between the NV center crystal directions and the applied magnetic field which change with respect to each crystal axis. There are four NV center axes that create four energy levels that are degenerate with the field in plane of the single crystal diamond. The frequency corresponding to the 0 to -1 transition is used, since this is lower frequency and energy and thus more likely to occur. Using this calculated NV center resonance frequency, the spinwave wavevector is calculated according to Eqn. \ref{eqn:1} and the $\lambda_\text{{res}}$ is taken to be equal to the associated wavelength.

\bibliography{OpticallydetectedFMRSuppMatCitations}